# Inferring cell-specific lncRNA regulation with single-cell RNA-sequencing data in the developing human neocortex

Meng Huang[1], Jiangtao Ma[2,3], Changzhou Long[1], Junpeng Zhang[3,*], Xiucai Ye[1,4,*], Tetsuya Sakurai[1,4]

[1]Department of Computer Science, University of Tsukuba, Tsukuba, 3058577, Japan, [2]Department of Automation, Xiamen University, Xiamen, 361005, China, [3]School of Engineering, Dali University, Dali, 671003, China, [4]Center for Artificial Intelligence Research, University of Tsukuba, Tsukuba, 3058577, Japan

[*]Corresponding author:
Junpeng Zhang. E-mail: zhangjunpeng411@gmail.com
Xiucai Ye. E-mail: yexiucai@cs.tsukuba.ac.jp

## Abstract

Long non-coding RNAs (lncRNAs) are important regulators to modulate gene expression and cell proliferation in the developing human brain. Previous methods mainly use bulk lncRNA and mRNA expression data to study lncRNA regulation. However, to analyze lncRNA regulation regarding individual cells, we focus on single-cell RNA-sequencing (scRNA-seq) data instead of bulk data. Recent advance in scRNA-seq has provided a way to investigate lncRNA regulation at single-cell level. We will propose a novel computational method, *CSlncR* (cell-specific lncRNA regulation), which combines putative lncRNA-mRNA binding information with scRNA-seq data including lncRNAs and mRNAs to identify cell-specific lncRNA-mRNA regulation networks at individual cells. To understand lncRNA regulation at different development stages, we apply *CSlncR* to the scRNA-seq data of human neocortex. Network analysis shows that the lncRNA regulation is unique in each cell from the different human neocortex development stages. The comparison results indicate that *CSlncR* is also an effective tool for predicting cell-specific lncRNA targets and clustering single cells, which helps understand cell-cell communication.

## 1. Introduction

Long non-coding RNAs (lncRNAs) are non-coding RNAs lacking protein coding capacity with transcript size >200 nucleotides (Elling *et al.*, 2016). The existing evidences have shown that lncRNAs have important roles in biological process (Pan *et al.*, 2019; Ponting *et al.*, 2009), neural development (Sauvageau *et al.*, 2013) and brain disorders (Qureshi *et al.* 2010; Mishra and Kumar 2021). In a cell, lncRNA can regulate gene expression causing messenger RNA (mRNA) degradation or post-transcriptional repression (Luo *et al.,* 2016), which have an important effect for fundamental cellular functions, such as cell proliferation, cell differentiation and cell death (Yao et al., 2019). Therefore, identifying lncRNA-mRNA regulatory networks is helpful for revealing functions and regulatory mechanisms of lncRNAs.

As each single-cell is special in its microenvironment, lncRNA regulation is also usually unique in each cell. Previous studies have shown that many computational methods are developed to explore lncRNA regulation (Liao et al., 2011; Guo et al., 2015; Du et al., 2017; Wu et al., 2016) in the groups of cells by using bulk RNA-sequencing data. This may conceal the heterogeneity of lncRNA regulation across individual cells. Single-cell RNA-sequencing (scRNA-seq) technologies revolutionize the throughput and resolution of bulk RNA sequencing in transcriptome studies (Diego *et al.*, 2014; Tal Nawy, 2014), which provides a way to investigate lncRNA regulation at the single-cell level.

To explore lncRNA regulation involved in cellular processes in the developing human neocortex, Liu et al. combined scRNA-seq with bulk tissue RNA-seq, which helps deeply profile lncRNA expression in each development stage of human neocortical (Liu *et al.*, 2016). They have found that there are many abundantly expressed lncRNAs in single cells. However, lncRNA regulation is not revealed in each single-cell from different development stages of human neocortex. To investigate the heterogeneity of lncRNA regulation among individual cells, it is crucial to infer cell-specific lncRNA regulation. This indicates that one lncRNA regulatory network contains all lncRNA regulations in one cell.

In this work, to explore cell-specific lncRNA-mRNA regulatory networks, we use the cell-specific network (*CSN*) method (Dai *et al.* 2019) in single-cell lncRNA-mRNA sequencing data. *CSN* can be used to infer $m$ cell-specific networks in scRNA-seq data including $m$ cells and $n$ genes. To calculate the association strength between gene-gene in a cell, we use the statistic (see Eq. (2) in "Material and methods") to perform a one-side hypothesis test. The null and alternative hypothesis are whether two genes are independent or associated in cell $k$. The gene-gene association exists if its statistic is larger than a significant level (e.g. 0.05). However, we can not use *CSN* directly to identify cell-specific lncRNA regulatory networks. Here, we proposed *CSlncR* (cell-specific lncRNA regulation) by extending *CSN* to infer cell-specific lncRNA regulation. Firstly, *CSN* mainly explore all types of gene-gene interactions in the scRNA-seq data. Given the single-cell lncRNA-mRNA sequencing data, we are aimed at inferring the interactions between lncRNAs and mRNAs, rather than all interaction types (mRNA-mRNA, lncRNA-lncRNA, and lncRNA-mRNA). Next, *CSN* identifies cell-specific networks without using prior knowledge, which is an unsupervised method. The putative lncRNA-mRNA binding information is incorporated into *CSlncR* as prior knowledge. The proposed *CSlncR* has been applied to the single-cell lncRNA-mRNA sequencing data across individual cells of different development stages in the developing human neocortex. Network analysis results show that lncRNA regulation is unique in each cell. The comparison analysis indicates that *CSlncR* is an effective method for predicting cell-specific lncRNA targets, inferring lncRNA regulation at individual cells in different human neocortex development stages and understanding cell-cell communication.

## 2. Materials and methods

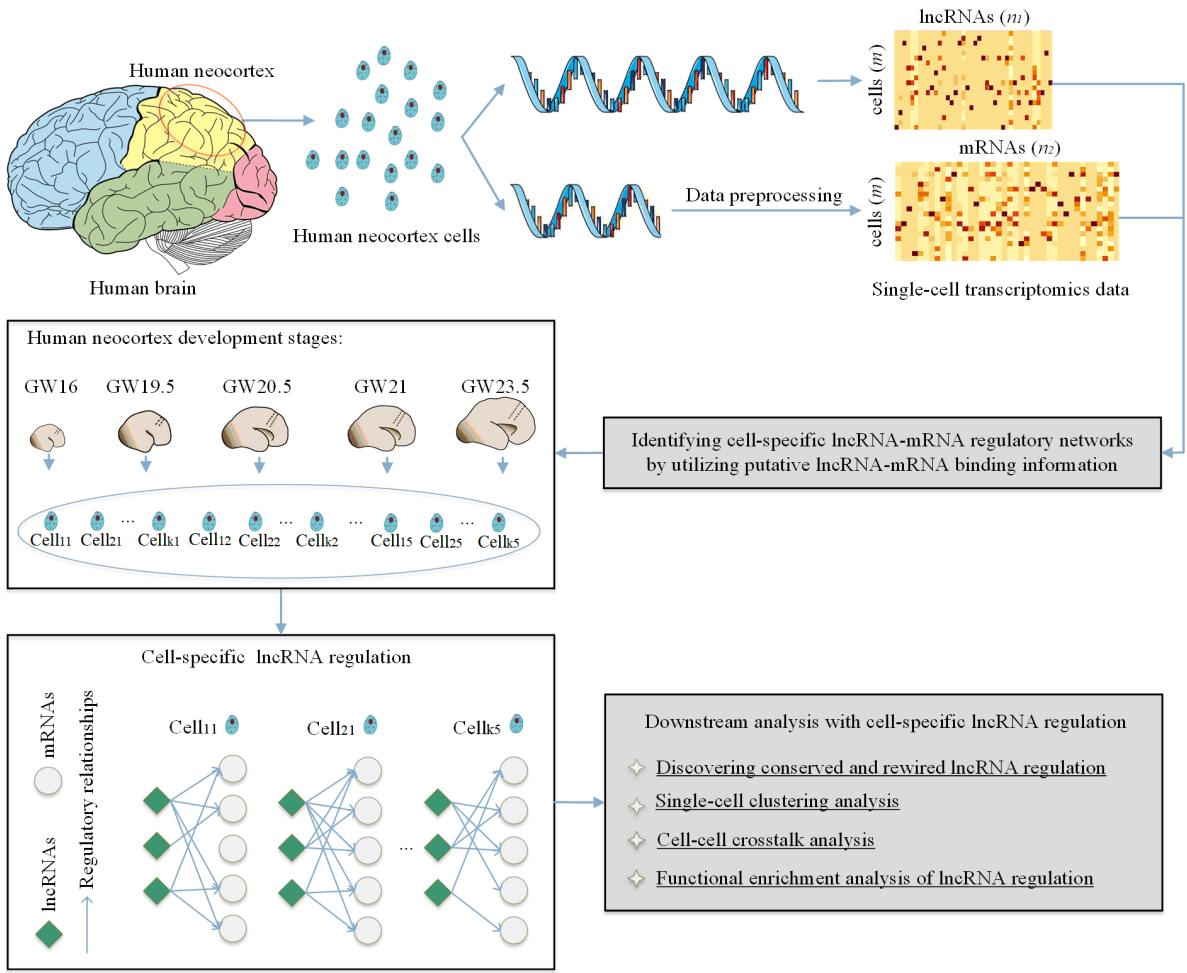

**Figure 1.** Workflow of *CSlncR*. Firstly, we preprocess 276 single-cells from the single-cell transcriptomics data in the developing human neocortex. Next, we identify $m$ cell-specific lncRNA-mRNA regulatory networks using putative lncRNA-mRNA binding information for $m$ cells (one lncRNA-mRNA regulatory

network with one cell). Finally, we conduct downstream analysis with identified $m$ cell-specific lncRNA-mRNA regulatory networks.

To investigate cell-specific lncRNA regulation in single-cell lncRNA-mRNA sequencing data, we propose the cell-specific lncRNA regulation (*CSlncR*). As shown in Figure 1, our method *CSlncR* includes three aspects: (1) obtaining and preprocessing single-cell transcriptomics data in the developing human neocortex; (2) identifying cell-specific lncRNA-mRNA regulatory networks by using putative binding information between lncRNAs and mRNAs; (3) performing downstream analysis in the identified cell-specific lncRNA regulation networks. In the following, we will describe three components of proposed *CSlncR* in detail.

## 2.1 Single-cell lncRNA-mRNA sequencing data in the developing human neocortex

We obtained the scRNA-seq data of human neocortex in the Gene Expression Omnibus database, where this data contain 276 cells and 13,007 genes with accession number GSE71315. Given the microdissected radial sections of the tissue at gestational weeks (GW) (Liu *et al.*, 2016), this dataset includes 5 development stages of human neocortex: GW16 (26 cells), GW19.5 (26 cells), GW20.5 (123 cells), GW21 (24 cells), and GW23.5 (77 cells). Here, we are only interested in two types: lncRNAs and mRNAs. Accordingly, we conducted a gene annotation using HGNC (HUGO Gene Nomenclature Committee, https://www.genenames.org/). Given the duplicate lncRNAs or mRNAs at each single-cell, we regarded the average expression values of same gene symbols as the final expression values. Besides, we also removed part of lncRNAs and mRNAs with constant expression values to reduce the nongenetic cell-to-cell variability (Golov *et al.*, 2016). Next, we performed the pre-processing using $log_2(x + 1)$ transformation to reduce the error produced by the technical noise in the scRNA-seq dataset. Finally, we have obtained 247 lncRNAs and 10,208 mRNAs of 276 single-cells in the developing human neocortex.

## 2.2 Identifying cell-specific lncRNA-mRNA regulatory networks by utilizing putative binding information between lncRNAs and mRNAs

To construct cell-specific lncRNA-mRNA regulatory networks in real human neocortex cells, we need to use an effective statistic to measure the association between lncRNAs and mRNAs. Given the robustness of *CSN* method for technical noises from single-cell sequencing technology, this statistic (Dai *et al.* 2019) is used in the proposed *CSlncR*. For each single-cell of human neocortex, this statistic helps *CSlncR* discovery lncRNA-mRNA regulation when building the regulatory network. Furthermore, we utilize the putative binding information between lncRNAs and mRNAs from NPInter v4.0 (Teng *et al.* 2020), ENCORI (Li *et al*., 2014) and LncRNA2Target (Cheng *et al*., 2019) as prior knowledges, which helps reduce computation between lncRNAs and mRNAs.

To evaluate the association between lncRNAs and mRNAs, we perform the statistical test for each putative lncRNA-mRNA pair ($lncR_u$ and $mR_v$) in a cell. To illustrate this statistic, we utilize the expression values of $lncR_u$ and $mR_u$ to draw the scatter diagram. As shown in Figure S1 (Supplementary Material 1), $u_k$ and $v_k$ represent expression values of $lncR_u$ and $mR_u$ in cell $k$, respectively. The neighbourhoods of $u_k$, $v_k$, and $(u_k, v_k)$ are denoted by the medium, light and dark grey box respectively. $m_v^{(k)}$, $m_u^{(k)}$, and $m_{uv}^{(k)}$ denote the number of points in the light, medium and dark grey boxes, respectively. Accordingly, the statistic $\rho_{uv}^{(k)}$ is defined as follows:

$$\rho_{uv}^{(k)} = \frac{m_{uv}^{(k)}}{m} - \frac{m_u^{(k)}}{m} \cdot \frac{m_v^{(k)}}{m}, \tag{1}$$

where $m$ is the number of human neocortex cells, $\frac{m_u^{(k)}}{m}$ and $\frac{m_v^{(k)}}{m}$ are the marginal probabilities of $lncR_u$ and $mR_v$, respectively, and $\frac{m_{uv}^{(k)}}{m}$ is the joint probability of $lncR_u$ and $mR_v$. Empirically, $\frac{m_u^{(k)}}{m} = \frac{m_v^{(k)}}{m} = 0.1$ according to the *CSN* method. Besides, we normalize the statistic $\rho_{uv}^{(k)}$ as follows:

$$\hat{\rho}_{uv}^{(k)} = \frac{\rho_{uv}^{(k)} - \mu_{uv}^{(k)}}{\sigma_{uv}^{(k)}} = \frac{\sqrt{m-1} \cdot (m \cdot m_{uv}^{(k)} - m_u^{(k)} m_v^{(k)})}{\sqrt{m_u^{(k)} m_v^{(k)} (m - m_u^{(k)})(m - m_v^{(k)})}}, \tag{2}$$

where $\mu_{uv}^{(k)} = 0$ and $\sigma_{uv}^{(k)} = \sqrt{\frac{m_u^{(k)} m_v^{(k)} (m - m_u^{(k)})(m - m_v^{(k)})}{m^4(m-1)}}$ are the mean value and standard deviation for the statistic $\rho_{uv}^{(k)}$, respectively. $\hat{\rho}_{uv}^{(k)}$ follows standard normal distribution ($\hat{\rho}_{uv}^{(k)} \sim N(0,1)$). Here, we obtain $p$ value of each $\rho_{uv}^{(k)}$ to measure the association significance between $lncR_u$ and $mR_v$. The smaller the $p$ value, the higher the association probability between $lncR_u$ and the $mR_v$ in cell $k$. Empirically, we set the cutoff of $p$ to 0.05.

In cell $k$, the corresponding smaller significance $p$ value (i.e. less than 0.05) indicates that there are an association between $lncR_u$ and $mR_v$. As we only focus on the lncRNA-mRNA regulatory networks in each human neocortex cell (276 single-cells), 276 cell-specific lncRNA-mRNA regulatory networks are kept at the end. Since the given human neocortex scRNA-seq dataset includes 5 development stages, there are 26, 26, 123, 24 and 77 cell-specific lncRNA-mRNA regulatory networks for GW16, GW19.5, GW20.5, GW21, and GW23.5, respectively. The bipartite graph is used to represent the lncRNA-mRNA regulatory network in each cell, in which lncRNAs or mRNAs are indicated by nodes, and the pointing relationship from a lncRNA to a mRNA is represented by edges.

## 2.3 Downstream analysis with cell-specific regulatory networks

To explore gene regulation at the network level, we focus on gene regulatory networks. In the previous step, the identified cell-specific lncRNA-mRNA regulatory networks help investigate lncRNA regulation in the developing human neocortex. For the proposed *CSlncR*, we perform the following downstream analysis to explore lncRNA regulation at different development stages of human neocortex: (1) discovering conserved and rewired lncRNA regulation, (2) single-cell clustering analysis, (3) cell-cell crosstalk analysis, and (4) functional enrichment analysis of lncRNA regulation.

### 2.3.1 Discovering conserved and rewired lncRNA regulation

The regulation for some lncRNAs is "on" in a cell or multiple cells whereas the regulation of some lncRNAs is "off" (Peng, Koirala, and Mo, 2017; Zhang, *et al.* 2019). To unveil the heterogeneity and commonality across different cells, we focus on this "on/off" state of lncRNA regulation. As the lncRNA regulation can be used to characterize each cell, we explore the conserved and rewired lncRNA regulation in each cell. These identified lncRNA regulations unveil the heterogeneity and commonality of cells. From the perspective of both lncRNA-mRNA regulatory network and hub lncRNAs, we discover that the conserved and rewired lncRNA regulation are very crucial. Previous studies (Hahn, and Kern 2005; Song, and Singh 2013) have shown that the essential nodes consists of nearly 20% of the nodes in a biological network. Accordingly, the top 20% of lncRNAs calculated by using node degrees are viewed as hub lncRNAs for each cell-specific lncRNA regulatory network. In general, we rank ~90% lncRNA-mRNA interactions or hub lncRNAs as a highly conservative level. For example, there are 26 single-cells for GW16 development stage of human neocortex. The lncRNA-mRNA interactions or hub lncRNAs that are always "on" in at least 23 (~ 90%) human neocortex cells for GW16 development stage are regarded as conserved interactions or hubs. Besides, the lncRNA-mRNA interactions or hub lncRNAs that are "on" in only one single-cell are viewed as rewired interactions or hubs. At the end, we identify conserved and rewired lncRNA-mRNA regulatory networks or hub lncRNAs, which helps explore the heterogeneity and similarity of lncRNA regulation across single-cells of human neocortex.

### 2.3.2 Single-cell clustering analysis

To understand tissue complexity (Kiselev, Andrews, and Hemberg 2019), we can cluster single cells of human neocortex scRNA-seq data using cell-cell similarity matrices represented by lncRNA-mRNA interactions or hub lncRNAs. Given the general similarity calculation (Zhang, *et al.* 2018), we obtain the similarity of interaction and hub lncRNA as follows:

$$S_{ij} = \frac{intersect(n_i, n_j)}{min(n_i, n_j)}, \tag{3}$$

where $n_i$ and $n_j$ represent the number of interactions or hub lncRNAs in the cell-specific lncRNA-mRNA regulatory networks of cells $i$ and $j$, respectively, $intersect(n_i, n_j)$ indicates the number of lncRNA-mRNA interactions or hub lncRNAs shared in the corresponding cell-specific lncRNA-mRNA regulatory networks.

To compare clustering performance with different similarity measures, we also obtain the similarity using the single-cell gene expression data below:

$$nor\_d_{ij} = \frac{d_{ij} - \min(d)}{max(d) - \min(d)} \tag{4}$$

where $d_{ij} = \sqrt{(e_{i1} - e_{j1})^2 + \cdots + (e_{it} - e_{jt})^2 + \cdots + (e_{ig} - e_{jg})^2}$, $d = (d_{ij}) \in \mathbb{R}^{m \times m}$, $e_{it}$ and $e_{jt}$ represent the gene ($t$) expression values in cells $i$ and $j$ respectively, $g$ is the number of mRNAs and lncRNAs, $m$ is the number of human neocortex single-cells, $nor\_d_{ij}$ is the normalized Euclidean distance between cells $i$ and $j$. After obtaining two similarity matrices corresponding to lncRNA-mRNA interactions and hub lncRNAs and one distance matrix $d$ about single-cell gene expression data, we perform single-cell hierarchical clustering analysis.

### 2.3.3 Cell-cell crosstalk analysis

The cell-cell crosstalk is an indirect communication between cells, which can influence gene expression patterns (Kaminska, *et al.* 2018), and the development and regeneration of the respiratory system as well (Zepp, and Morrisey, 2019). Here, we use the similarity matrices to conduct the cell-cell crosstalk analysis. Previous studies (Zhang, *et al.* 2020; Shi, *et al.* 2017) have shown that lncRNA regulation plays a key role in cell signaling pathways of human brain. In each cell pair, the higher the similarity, the more general the shared cell signaling pathways between two cells. This also means that the cell pair is more likely to have signals to communicate with each other (crosstalk). To measure the crosstalk relationship between two cells, we set the cutoff to the median value of all cell-cell similarities based on the interaction or hub lncRNA similarity matrix. If the similarity value between cells is larger than this median value, there is a crosstalk relationship between cell $i$ and cell $j$. To obtain a cell-cell crosstalk network, we assemble the crosstalk relationship between cells based on lncRNA-mRNA interactions or hub lncRNAs.

### 2.3.4 Functional enrichment analysis of lncRNA regulation

To evaluate the identified cell-specific lncRNA-mRNA regulatory networks, we perform functional enrichment analysis of lncRNA regulation at the network level. Moreover, we utilize the experimentally validated lncRNA-mRNA binding information from a well-known database named NPInter v4.0 (Teng *et al.* 2020), LncTarD (Zhao *et al.*, 2020) and LncRNA2Target (Cheng *et al.*, 2019) for validation. Since the human neocortex cells are highly correlated with autism spectrum disorder (ASD), we group a set of mRNAs and lncRNAs associated with ASD to explore ASD-related lncRNA regulation. We obtained these ASD-related lncRNAs and mRNAs in the previous ASD-related studies (Cogill, *et al.* 2018; Wang, *et al.* 2015; Ziats, and Rennert, 2013) and SFARI Gene Database 2.0 (Abrahams, *et al.* 2013), which is an evolving database for the available genes and variants related to ASD in autism susceptibility.

To uncover potential biological functions of identified lncRNA-mRNA regulatory networks in each development stage of human neocortex, we use the clusterProfiler R package (Yu, *et al.* 2012) to perform Gene Ontology (GO) (Ashburner, *et al.* 2000), Kyoto Encyclopedia of Genes and Genomes (KEGG) (Kanehisa, and Goto, 2000), Reactome (Fabregat, *et al.* 2018), Hallmark (Subramanian, *et al.* 2005), Cell marker (Zhang, *et al.* 2019) enrichment analysis. If the $p$ value of terms adjusted by Benjamini–Hochberg method is less than 0.05, we regard them as significant terms. To evaluate whether the lncRNAs and mRNAs in each human neocortex development stage are significantly enriched in ASD or not, we use a hyper-geometric test to conduct ASD enrichment analysis. For each human neocortex development stage, we can obtain the significance $p$ value enriched in ASD as follows:

$$p\_value = 1 - \sum_{t=0}^{r-1} \frac{\binom{S}{t}\binom{N-S}{M-t}}{\binom{N}{M}} , \qquad (4)$$

where $N$ indicates the number of lncRNAs and mRNAs in the human neocortex scRNA-seq dataset. $S$ represents the number of ASD-related genes in this dataset, $M$ and $r$ are the number of genes and ASD-related genes respectively in each human neocortex development stage. Here, the cutoff of $p$ value is set to 0.05.

## 3. Results and discussion

### 3.1 The lncRNA regulation is unique in each cell of human neocortex

To investigate the cell-specific lncRNA regulation in human neocortex, we propose *CSlncR* to infer the lncRNA-mRNA interactions of interest. Using the proposed *CSlncR*, we have obtained 276 cell-specific lncRNA-mRNA regulatory networks for the 276 human neocortex cells. Since there are 5 development stages including GW16, GW19.5, GW20.5, GW21 and GW23.5 in the human neocortex scRNA-seq dataset, we explore the uniqueness of each human neocortex cell based on cell-specific lncRNA-mRNA interactions and hub lncRNAs by showing the results of investigation for each human neocortex development stage.

Firstly, we have explored four perspectives according to the identified cell-specific lncRNA-mRNA regulatory networks and hub lncRNAs: (1) the number of predicted lncRNA-mRNA interactions, (2) the percentage of validated lncRNA-mRNA interactions, (3) the percentage of ASD-related lncRNA-mRNA interactions, (4) the percentage of ASD-related hub lncRNAs. As shown in Figure 2A-B, we find that the number of predicted lncRNA-mRNA and the percentage of validated lncRNA-mRNA are different in each of 26 cells from GW16 development stage of human neocortex. Moreover, the percentage of ASD-related lncRNA-mRNA interactions and hub lncRNAs is also different across each cell in Figure 2C-D. This result shows that the lncRNA regulation is different in each cell. Similarly, we also verify lncRNA regulation of cells in other development stages (GW19.5, GW20.5, GW21 and GW23.5). As shown in Figure S2, S3, S4, S5 (Supplementary Material 1), the lncRNA regulation is also unique in each cell of the other four development stages. Furthermore, we investigate the conserved and rewired lncRNA-mRNA interaction in each human neocortex development stage (GW16, GW19.5, GW20.5, GW21 and GW23.5). For example, in GW16 development stage, the percentage of the conserved and rewired interactions is 0.45% (37 out of 8,170) and 21.49% (1756 out of 8,170), respectively. This results indicate that a part of lncRNA-mRNA interactions tend to be rewired across human neocortex cells in GW16. For other human neocortex development stages (GW19.5, GW20.5, GW21 and

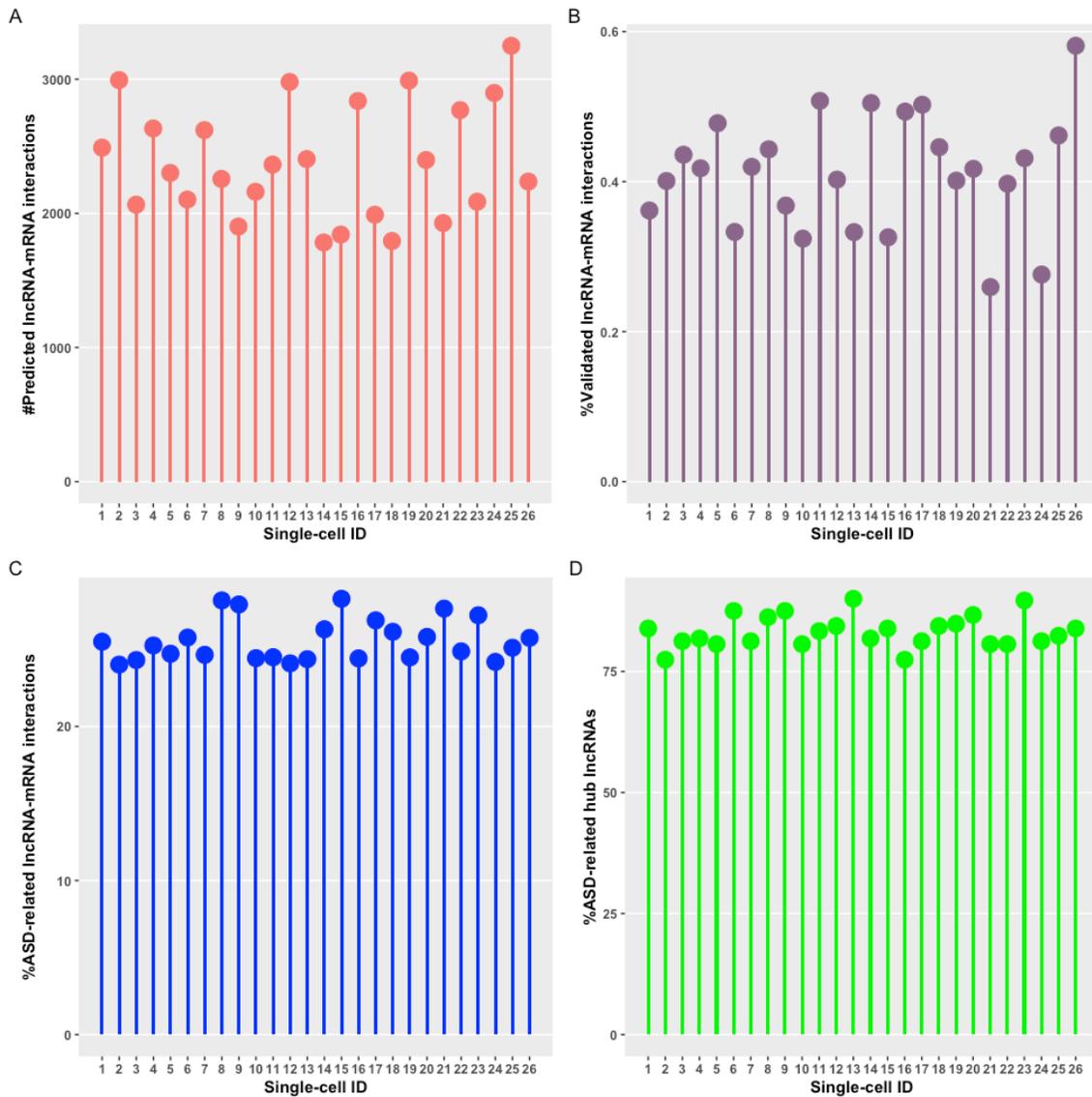

**Figure 2.** Cell-specific lncRNA-mRNA interactions and hub lncRNAs in GW16 development stage of human neocortex. (A) Number of predicted lncRNA-mRNA interactions. (B) Percentage of validated lncRNA-mRNA interactions. (C) Percentage of ASD-related lncRNA-mRNA interactions. (D) Percentage of ASD-related hub lncRNAs.

GW23.5), we can still obtain similar findings. In Supplementary Material 2, we list all detailed information of the conserved and rewired lncRNA-mRNA interactions. According to the similarity of the lncRNA-mRNA interactions from cell-specific regulatory networks, the similarity range between cells is [0.30, 0.77] in GW16 development stage. As shown in Figure 3(A), the cell similarity is less than 1 between any pair of 26 human neocortex cells in GW16. The cell similarity of other human neocortex development stages is shown in Figure S6(A), S7(A), S8(A) and S9(A) (Supplementary Material 1). Besides, we also investigate the conserved and rewired hub lncRNAs in each human neocortex development stages. For example, in GW16 development stage, the percentage of the conserved and rewired hub lncRNAs is 21.87% (14 out of 64) and 17.18% (11 out of 64), respectively. This indicates that nearly half of hub lncRNAs usually are conserved across human neocortex cells in GW16. In other development stages, we can obtain similar findings. In terms of the similarity of hub lncRNAs from the corresponding regulatory networks, the similarity range between cells is [0.58, 0.90] in GW16. As shown in Figure 3(B), the cell similarity is less than 1 between any pair of 26 human neocortex cells in GW16. In other human neocortex development stages, the cell similarity is shown in Figure S6(B), S7(B), S8(B) and S9(B) (Supplementary Material 1). In Supplementary Material 2, we list all detailed information about conserved and rewired hub lncRNAs.

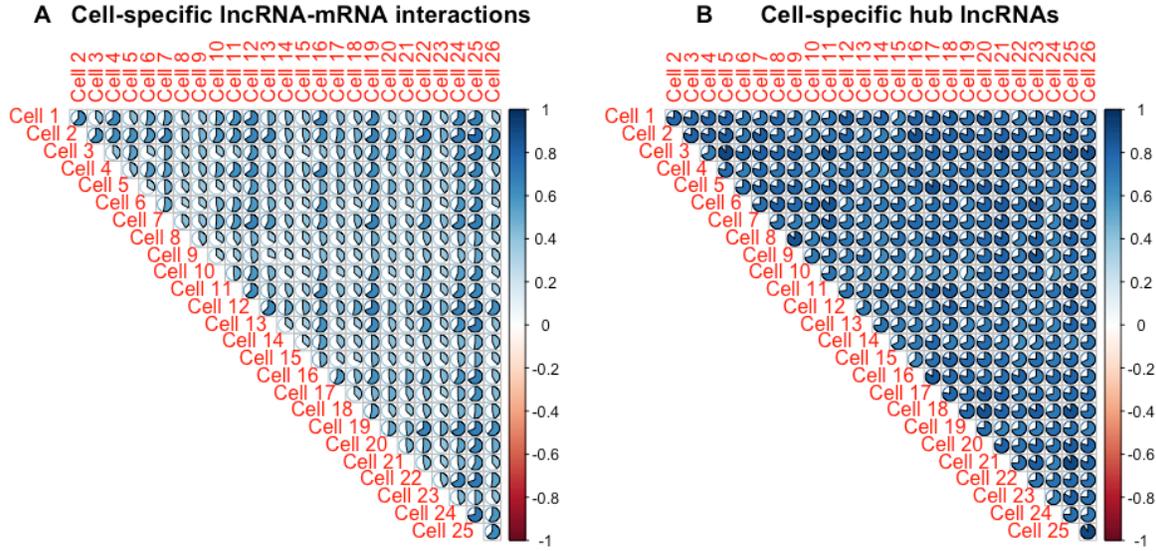

**Figure 3.** Single-cell similarity plot with 26 cells in GW16 development stage of human neocortex. (A) Similarity plot in terms of cell-specific lncRNA-mRNA interactions. (B) Similarity plot in terms of cell-specific hub lncRNAs. Colored areas indicate higher similarity between single cells.

In summary, for the cell-specific lncRNA-mRNA regulatory networks and hub lncRNAs, we find that there are always the difference of lncRNA regulation between any pair of cells in different human neocortex development stages. Accordingly, this also demonstrates the uniqueness of lncRNA regulation in each cell.

## 3.2 The cell type-specific regulation of lncRNA biomarkers across human neocortex single-cells

To further explore the lncRNA regulation across single-cells in human neocortex, we perform a case study to infer cell type-specific regulation for 18 identified cell type-specific lncRNA biomarkers (Liu, et al. 2016). In the previous study (Liu, et al. 2016), there are 7 known cell types related to 18 lncRNA biomarkers: endothelia (*LINC00339, TRIM52-AS1*), radial glia (*LINC00943, MAGI2-AS3, RUSC1-AS1*), dividing radial glia (*THAP9-AS1*), intermediate progenitors (*DGCR11*), newborn neurons (*INHBA-AS1, MYT1L-AS1, KIF9-AS1*), maturing excitatory (*MIR137HG*, PWAR6, SIK3-IT1, NAV2-AS3, DAPK1-IT1), inhibitory interneurons (*DLX6-AS1, SOX2-OT, MEG3*). These biomarkers are very important for cell cycle, cell proliferation and other key biological function (Liu, *et al.* 2016). Previous studies (Li, and Guo, 2020; Cogill, *et al.* 2018; Wang, *et al.* 2015) have shown that these cell type-specific lncRNAs biomarkers is in association with Autism Spectrum Disorder (ASD). Hence, we focus on the regulation of these cell type-specific lncRNAs in human neocortex.

To measure the significant difference in the regulation of cell type-specific lncRNAs between each pair of human neocortex cells, we compare the distributions of predicted target numbers and the distributions of ASD-related target percentages about cell type-specific lncRNAs in each human neocortex cell by using a two-sample Kolmogorov–Smirnov (KS) test (Conover, 1971). This non-parametric KS test can be used to assess whether the distribution of the predicted target numbers, or the ASD-related target percentages about cell type-specific lncRNAs in one human neocortex cell is significantly shifted compared with the distribution in another human neocortex cell. To obtain these distributions in each human neocortex cell, we calculate the predicted target numbers and the ASD-related target percentages about cell type-specific lncRNAs, respectively. As shown in Figure 4A-B, in the case of predicted targets and ASD-related targets, the regulations of cell type-specific lncRNAs between most of pairs of 26 human neocortex cell are significantly different (p value < 0.001). These results show that the regulation of cell type-specific lncRNAs is likely to be cell-specific in GW16 development stage. Similarly, we also obtain the same findings for the GW19.5, GW20.5, GW21 and GW23.5 development stages of human neocortex in Figure S10, S11, S12 and S13 (Supplementary Material 1). As shown in Figure S14(A), we find that the number of conserved targets of cell type-specific lncRNA regulation is less than the number of rewired targets of them. These difference indicates that the dominant cell type-specific lncRNA regulation type across human neocortex cells may be rewired lncRNA regulation in GW16 development stage. Similarly, we also find that the dominant biomarkers-related lncRNA regulation of the remaining development stages may be rewired in Figure S14(B, C, D, E). All detailed information about conserved and rewired targets related with cell type-specific lncRNAs biomarkers can be seen in Supplementary Material 3.

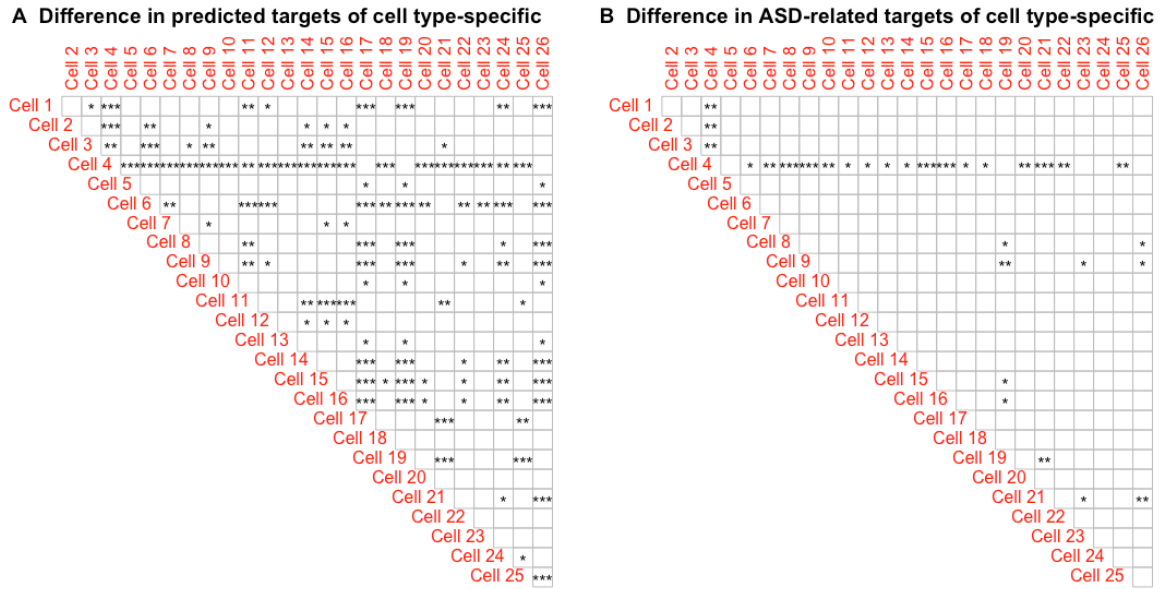

**Figure 4.** The cell type-specific lncRNA regulation between any pair of cells in GW16 development stage. (A) Difference in predicted targets of cell type-specific lncRNA. (B) Difference in ASD-related targets of cell type-specific lncRNA. Empty square shapes denote $p$ values > 0.05, * denotes $p$ value < 0.05, ** denotes $p$ value < 0.01 and *** denotes $p$ value < 0.001.

Generally, lncRNAs regulate target genes to perform a specific biological function. Here, we find that the rewired lncRNA regulation is likely to be the dominant regulation of lncRNA biomarkers in each human neocortex development stage. Therefore, we implement enrichment analysis of rewired lncRNA-mRNA regulatory interactions related with cell type-specific lncRNAs in each human neocortex development stage. The rewired lncRNA-mRNA regulatory networks in each human neocortex development stage are significantly enriched in at least four terms of Gene Ontology (GO), Kyoto Encyclopedia of Genes and Genomes Pathway (KEGG), Reactome, Hallmark or Cell marker in Table S1 (Supplementary Material 1). Besides, we find that the rewired lncRNA-mRNA regulatory networks associated with cell type-specific are enriched in ASD for GW19.5 and GW21 development stages. In more detail, many significant terms including the GO biological process "mRNA catabolic process" (Wang, *et al.* 2020), KEGG pathway "mRNA surveillance pathway" (Addington, *et al.* 2011), Reactome pathway "mRNA Splicing" (Parras, *et al.* 2018), and Hallmark "HALLMARK_MYC_TARGETS_V1" (Jin, *et al.* 2018) are closely associated with Autism spectrum disorder (ASD). These results show that the rewired lncRNA-mRNA regulatory networks about cell type-specific lncRNAs are functional across cells in each human neocortex development stage. All detailed enrichment analysis results for the rewired lncRNA-mRNA can be seen in Supplementary Material 4.

### 3.3 CSlncR is effective in predicting cell-specific lncRNA targets for different human neocortex development stages

To assess the effectiveness of CSlncR, we compare CSlncR with the other methods (CSlncR without using prior knowledges, Random method, LncRNA2Target (Cheng, et al., 2019) and NPInterPrediction (Teng et al., 2020)) according to the percentage of validated lncRNA-mRNA interactions across single-cells in each human neocortex development stage. As shown in Figure 5, we compare the results of different methods in GW16 development stage. In Figure 5(A), we compare CSlncR and its variant without utilizing prior knowledge to explore the effectiveness prior knowledge for the accuracy of lncRNA target prediction. As for as CSlncR and its variant without using prior knowledge, the validated average percentage of lncRNA-mRNA interactions across single-cells is 0.412% and 0.025%, respectively. To illustrate the advantages of CSlncR using prior knowledge compared with its variant without using prior knowledge for validated lncRNA-mRNA interactions, we use a paired t-test. Accordingly, the validated percentage of lncRNA-mRNA interactions across cells in GW16 by using prior knowledge for CSlncR is larger than that of its variant without using prior knowledge at a significant level ($p$ value = 2.20E-16). This indicates that the prior knowledge helps enhance the accuracy of lncRNA target prediction. Besides, we further compare CSlncR with a random method in Figure 5(B). In the identified cell-specific lncRNA-mRNA regulatory networks for GW16 development stage, we use the lncRNA-mRNA interactions generated by the random method to obtain the validated average percentage of lncRNA-mRNA interactions. For the random method, the validated average percentage of lncRNA-mRNA interactions across cells is 0.019%. The results using a paired t-test show that the validated percentage of lncRNA-mRNA interactions across cells by CSlncR is larger than that of the random method at a significant level ($p$ value = 2.20E-16). Furthermore, we compare CSlncR with LncRNA2Target, which is the most complete lncRNA-Target relationship database. For LncRNA2Target, the validated average percentage of lncRNA-mRNA interactions across cells is 0.339% in Figure 5(C). These results using a paired t-test indicate that the validated percentage of lncRNA-mRNA interactions across cells by CSlncR is also larger than that of LncRNA2Target at a significant level ($p$ value = 5.36E-05). Finally, we compare CSlncR with NPInterPrediction in Figure 5(D). For NPInterPrediction, the validated average percentage of lncRNA-mRNA interactions across cells is 0.012%. The result using a paired t-test indicates that the validated percentage of lncRNA-mRNA interactions across cells by CTSlncR is also larger than that of NPInterPrediction at a significant level ($p$ value = 2.20E-16). Similarly, we also compare CSlncR with other methods for GW19.5, GW20.5, GW21 and GW23.5 development stages. As shown in Figure S15, S16, S17 and S18 (Supplementary Material 1), we find that the validated percentage of lncRNA-mRNA interactions across cells by CTSlncR is also larger than that of other methods. In summary, these comparison results show that CSlncR is effective in predicting cell-specific lncRNA targets for different human neocortex development stages.

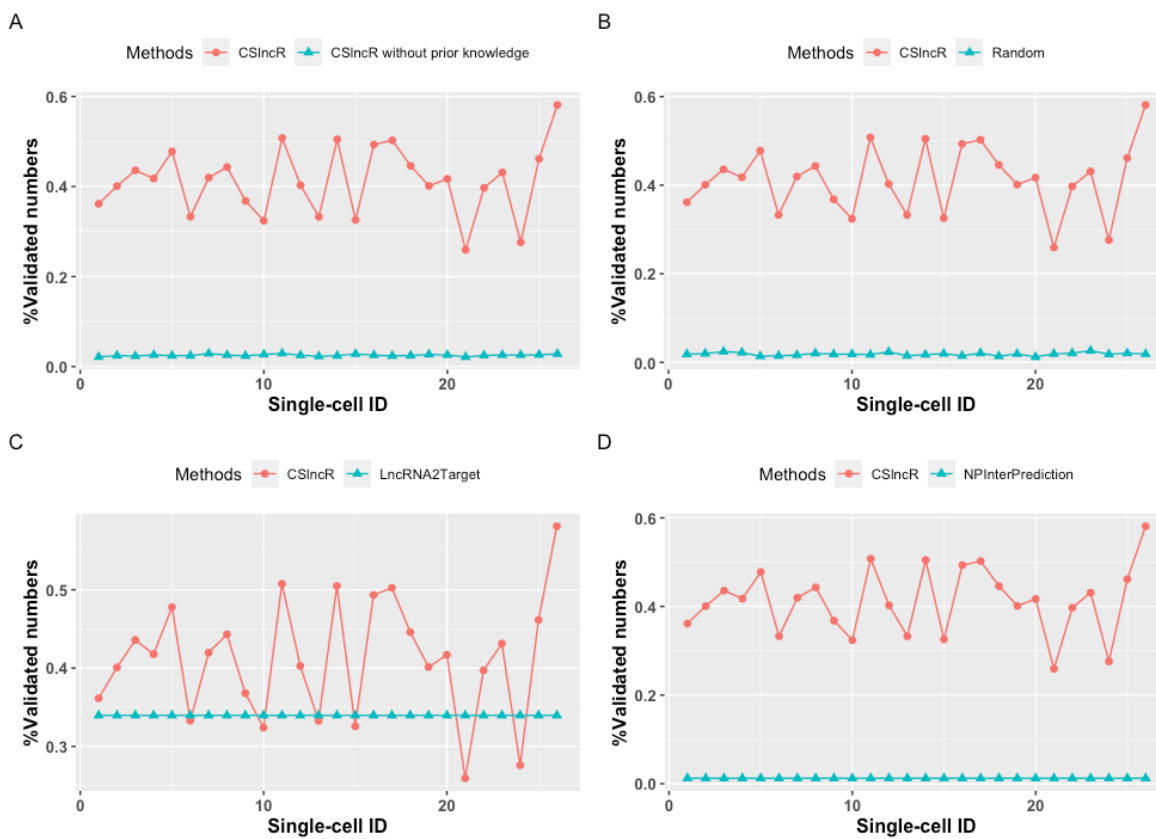

**Figure 5.** Comparison in terms of the percentage of validated lncRNA-mRNA interactions in GW16 development stage. (A) Comparison results between CSlncR (with prior knowledges) and CSlncR without prior knowledges. (B) Comparison results between CSlncR and Random method. (C) Comparison results between CSlncR and LncRNA2Target. (D) Comparison results between CSlncR and NPInterPrediction.

### 3.4 CSlncR provides a novel strategy to cluster single-cells

Previous clustering methods mainly use the scRNA-seq expression data to perform cluster analysis, which helps cluster single-cells. Here, we use the interaction and hub lncRNA similarity generated by CSlncR to cluster single-cells. As a result, we compare the clustering results using interaction similarity and hub lncRNA similarity with the result of the clustering using the Euclidean distance (normalized to the range of [0, 1]) based on the scRNA-seq expression data.

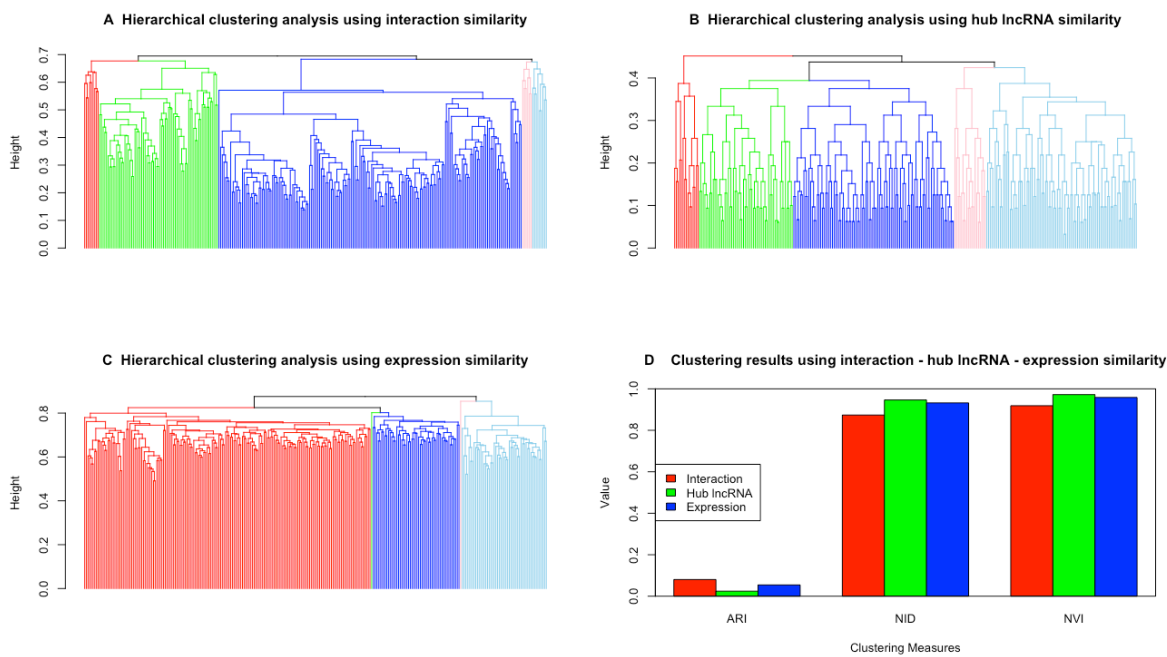

**Figure 6.** Hierarchical cluster analysis of the 276 single-cells in 5 human neocortex development stages. (A) Hierarchical cluster analysis by using interaction similarity. (B) Hierarchical cluster analysis by using hub lncRNA similarity. (C) Hierarchical cluster analysis by using expression similarity. Each color denotes a cluster. (D) Clustering results using interaction, hub lncRNA, and expression similarity, respectively.

As shown in Figure 6, the hierarchical clustering is used to cluster single-cells based on interaction similarity (Figure 6A) and hub lncRNA similarity (Figure 6B) respectively, compared with the clustering results based Euclidean distance (Figure 6C). These clustering results are different because of the used similarity measure and distance measure. However, the clustering results using interaction similarity and hub lncRNA similarity show 5 distinct clusters corresponding to 5 human neocortex development stages, whereas the clustering result using the gene expression data does not indicate clear 5 clusters. Furthermore, we perform quantitative analysis for the hierarchical clustering results by using 3 clustering measures including Adjusted Rand Index (ARI), Normalized Information Distance (NID), and Normalized Variation of Information (NVI). As shown in Figure 6 (D), the clustering accuracy according to the interaction similarity and hub lncRNA similarity have outperformed that of gene expression. As for these results, previous studies have shown that gene regulatory networks help represent the cell status and the biological process compared with gene expressions (Liu, *et al.* 2017; Yang, *et al.* 2018).

In the future, the wet-lab experiments can be further applied to validating our clustering analysis results. To help biological researcher discover potential cell clusters indicating novel cell subtypes, CSlncR provides a novel strategy to cluster these single-cells.

### 3.5 CSlncR is helpful for understanding cell-cell crosstalk

Previous studies (Sharma, et al. 2013; Charles, *et al.* 2011) have shown that cell-cell communication or crosstalk is very important for multicellular organisms (i.e. human). This communication mechanism makes multiple cells communicate and cooperate, which helps maintain important life activity. Here, the crosstalk relationship between cell $i$ and cell $j$ depends on whether the similarity value between cell $i$ and cell $j$ is larger than the median similarity value or not. The cell-cell crosstalk relationships can be used to produce the crosstalk networks between cells by using interaction similarity or hub lncRNA similarity. In the interaction and hub lncRNA similarity matrices, we can obtain two crosstalk networks between cells in each human neocortex development stage. All detailed information is shown in Supplementary Material 5.

Those identified cell-cell crosstalk networks show the communication frequency between cells. These frequently communicated cells can be viewed as hub cells or active cells. In detail, the top 20% of cells according to the rank of node degrees is regarded as hub cells in each cell-cell crosstalk network from 5 human neocortex development stages. In Table S2, S3, S4, S5 and S6 (Supplementary Material 1), these hub cells may be regarded as key factors to connect different subtypes of human neocortex single-cells from different development stages (GW16, GW19.5, GW20.5, GW21 and GW23.5). Besides, we can explore which cells are prone to gather a single module in the process of communication. For the identified cell-cell crosstalk networks, we use the Markov Clustering Algorithm (Enright, *et al.* 2002) to identify cell-cell crosstalk modules, where the number of human neocortex single-cells is at least 3 for each module. In Table S7, S8, S9, S10 and S11 (Supplementary Material 1), we discover that most of the human neocortex single-cells are gathered into one module to communicate with each other in each human neocortex development stage. Since the used 276 human neocortex single-cells may be phenotypically identical, most of them are prone to gather a module in cell-cell crosstalk.

## 4. Conclusions

As we all know, lncRNA regulation plays an important role in the key biological processes including RNA silencing, transcriptional regulation of gene expression, ASD transcriptional regulation, cellular functions, signaling pathways, and neurodegenerative and neuropsychiatric diseases. In the previous studies (Shi, *et al.* 2016; Statello, *et al.* 2021), lncRNA regulation is specific to conditions, which indicates that the lncRNA regulation may be cell-specific even for these phenotypically identical single-cells. Given the scRNA-seq technology, we enable to explore lncRNA regulation at individual cells. In this work, to develop cell-specific lncRNA-mRNA regulatory networks at single-cell level, we propose a novel method, CSlncR, which helps infer cell-specific lncRNA regulation in the developing human neocortex by using the networks. Since we identify these cell-specific lncRNA-mRNA regulatory networks based on scRNA-seq data with using prior knowledge, the proposed CSlncR can be viewed as a supervised method.

In real applications, there are some limitations for the proposed CSlncR. In the future, we consider improving CSlncR from the following aspects. Firstly, we plan to improve the accuracy of predicted cell-specific lncRNA-mRNA regulatory networks. The comprehensive lncRNA-mRNA binding information can be incorporated into CSlncR as prior knowledge. Secondly, we identify all cell-specific lncRNA-mRNA networks based on correlation relationships. Actually, we consider to unveil lncRNA causal regulation in single cell. These cell-specific lncRNA regulatory networks can be identified by using the causal relationships instead of correlation relationships. Finally, we need to consider the type of lncRNA regulation. In the process of lncRNA regulation, there are generally lncRNA-directed regulation where lncRNAs directly regulates the expression of mRNAs and lncRNA-indirected regulation where lncRNAs are viewed as mediators to be related with gene regulation. In this work, we only consider lncRNA-directed regulation. In the competing endogenous RNA (ceRNA) hypothesis (Salmena *et al.* 2011), lncRNAs can be viewed as mediators to be related with the crosstalk between different RNA transcripts including mRNAs, microRNAs, circular RNAs and transcribed pseudogenes. Furthermore, the cell-specific lncRNA sponge interaction networks can also be inferred in future.

Although there are some limitations, CSlncR still helps investigate the heterogeneity of lncRNA regulation in each single-cell from human neocortex. Especially, we apply CSlncR to the study of brain disorders (e.g. ASD) (Liu, *et al.* 2016; Cogill, *et al.* 2018), where a few cells could be profiled. At single-cell level, CSlncR can be an effective method, which helps study the noncoding RNA (e.g. lncRNA) for the biologists.

## Data availability statement

The source code used to replicate all our analyses, including all real data, is available at the following link: https://github.com/linxi159/CSlncR

## Competing interests

The authors declare that they have no known competing financial interests or personal relationships that could have appeared to influence the work reported in this paper.

## Supplementary data

Supplementary data are available online at the following link:

https://github.com/linxi159/CSlncR/tree/main/Supplementary_file

## Acknowledgments

We thank the editor and reviewers for their help and comments during the preparation of the manuscript. We acknowledge the Support for Pioneering Research Initiated by the Next Generation (SPRING) in Japan and the NCBI Gene Expression Omnibus database for providing their platforms and contributors for uploading their meaningful datasets

## Funding

This work was supported in part by JST SPRING (Grant Number JPMJSP2124).